\documentclass[preprint,amsmath,amssymb,superscriptaddress,showpacs]{revtex4}
\usepackage{graphicx}
\begin{document}
\renewcommand{\thesection}{\arabic{section}}
\renewcommand{\thesubsection}{\arabic{section}.\arabic{subsection}}
\renewcommand{\thefigure}{\arabic{figure}}
\baselineskip=0.7cm

\title{Effect of a gap opening on the conductance of graphene superlattices}
\author{M. Esmailpour}
\affiliation{Department of physics, Razi University, Kermanshah, Iran}
\author{A. Esmailpour}
\affiliation{Department of physics, Shahid Rajaei University, Lavizan, Tehran 16788, Iran}
\affiliation{School of Physics, Institute for Fundamental Sciences, (IPM) 19395-5531 Tehran, Iran}
\author{Reza Asgari~\footnote{Corresponding author: Tel: +98 21 22280692; fax: +98 21 22280415.\\ E-mail address: asgari@theory.ipm.ac.ir, }}
\affiliation{School of Physics, Institute for Fundamental Sciences, (IPM) 19395-5531 Tehran, Iran}
\author{M. Elahi}
\affiliation{Department of physics, Razi University, Kermanshah, Iran}
\author{M. R. Rahimi Tabar}
\affiliation{Department of Physics, Sharif University of Technology, 11365-9161, Tehran, Iran}
\affiliation{Institute of Physics, Carl von Ossietzky University, D-26111 Oldenburg, Germany}

\begin{abstract}

The electronic transmission and conductance of a gapped graphene superlattice
were calculated by means of the transfer-matrix method.
The system that we study consists of a sequence of electron-doped graphene as
wells and hole-doped graphene as barriers. We show that the transmission
probability approaches unity at some critical value of the gap. We also find
that there is a domain around the critical gap value for which the conductance
of the system attains its maximum value.
\end{abstract}

\maketitle

\section{Introduction}

Graphene, a single atomic layer of crystalline carbon on the
honeycomb lattice that consists of two interpenetrating
triangular sublattices A and B, has opened up a new field for
fundamental studies and applications
~\cite{novoselov1,novoselov2,novoselov3,novoselov4,geim}. Peculiar
electronic properties of graphene give rise to the possibility of overcoming
the limitations of the silicon-based electronic ~\cite{carbon}. The electronic
spectrum in graphene contains two zero energy at $K^+$ and $K^-$
points of the Brillouin zone, which are called the valleys or Dirac
points. The massless Dirac-like carriers in graphene
have almost semi-ballistic transport behavior with small resistance,
due to the suppression of the back-scattering process ~\cite{thermal}. The
mobility of the carriers in graphene is quite high
\cite{morozov1,morozov2,morozov3,morozov4}, and is much higher than the
electron mobility revealed in the semiconductor hetrostructures
\cite{eng1,eng2}.

In graphene sheets the type of particle (electrons or holes) and the density
of the carriers can be controlled by tuning a gate bias voltage
\cite{novoselov2,Zhang1,Nilsson}. In a gapless graphene electrical conduction
cannot be switched off by using the control voltages \cite{Katsnelson3}, which
is essential for the operation of conventional transistors. One can
overcome such difficulties by generating a gap in the electronic spectrum.
The band gap is a measure of the threshold voltage
and the on-off ratio of the field effect transistors \cite{FET1,FET2}.
Therefore, it is
essential to induce a band gap at the Dirac points in order to control the
transport of the carriers and integrating graphene into the semiconductor
technology. Consequently, band gap engineering in graphene
is a current topic of much interest with fundamental and applied
significance \cite{gap}.

In the literature several routes have been
proposed and applied to induce and to control a gap in graphene. One of them
is using quantum confined geometries, such as quantum dots or
nanoribbons
\cite{gap_ribbon1,gap_ribbon2,gap_ribbon3,gap_ribbon4,gap_ribbon5}. It has
been shown that the gap values
increase by decreasing of the nanoribbon width. An alternative way is
spin-orbit coupling whose origin is due to both intrinsic spin-orbit
interactions, or the Rashba interaction
\cite{gap_spin1,gap_spin2,gap_spin3,gap_spin4}. Yet another method to generate
a gap in graphene sheets is through an inversion symmetry breaking of the
sublattice, when the number of the electrons on A and B atoms are different
\cite{gapsub1,gapsub2,gapsub3,gapsub4}, e.g., graphene placement on proper
substrates
\cite{dop_lanzara1,dop_lanzara2,lanzara1,lanzara2,eva,gruneis1,gruneis2,giovannetti}.

Graphene superlattices, on the other hand, may be fabricated by adsorbing
adatoms on graphene surface by positioning and
aligning impurities with scanning tunneling microscopy \cite{eigler}, or
by applying a local top gate voltage to graphene \cite{huard}.
The transition of hitting massless particles in a clean \cite{bai} or
disordered \cite{nima} graphene-based superlattice structure has been studied.
It is shown that the conductivity of the system depends on the superlattice
structural parameters. Furthermore,
the superlattice structure of the graphene nanoribbons
was recently studied by using first-principles density-functional theory
calculations \cite{sevincli}. These calculations
showed that the magnetic ground state of the constituent ribbons, the
symmetry of the junction, and their functionalization by adatoms represent
structural parameters to the electronic and magnetic
properties of such structures. It would, therefore, be worthwhile to
investigate how the conductance of graphene superlattice junctions are
affected by a gap opening at the Dirac points.

In this paper we consider the sublattice symmetry-breaking mechanism due to
the fact that the densities of the particles associated with the on-site energy
for A and B sublattices are different or, equivalently, consider the intrinsic
spin-orbit interaction for a gap opening in a clean graphene superlattice. We
investigate the transmission probability of the Dirac fermions through the
system. In the cases of a single barrier and double barrier,
exact analytical analysis are carried out for calculating the transmission
probability. In addition, we show that the group delay time is different from
the dwell time for a system that consists of a gap opening.

The rest of the paper is organized as follows. The theory and method are
discussed in Section 2. The numerical results and discussions are given in
Sec. 3. A brief summary is given in Sec. 4.

\section{The Superlattice Model}

\begin{figure}[ht]
\begin{center}
\includegraphics[width=0.7\linewidth]{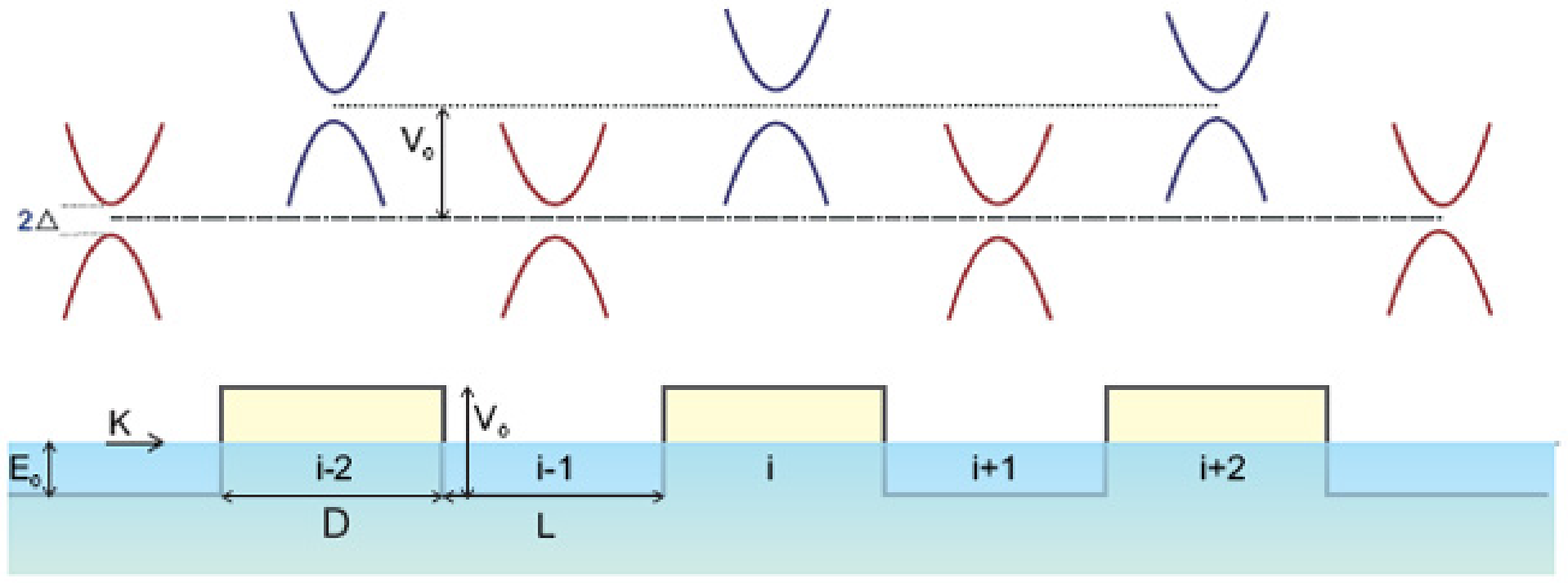}
\caption{Model of gapped graphene superlattices.}
\label{fig:t1}
\end{center}
\end{figure}

We consider a graphene with a peculiar gap opening due to the sublattice
symmetry breaking, where the 2D massive Dirac fermions at low energy is
described by noninteracting Hamiltonian \cite{alireza}
$\widehat{H}= \hbar v_{\rm F} \sigma \cdot {\bf k}+m v_{\rm F}^2 \sigma_z
$. There are two eigenvalues $\pm E_k$, where $E_k=\sqrt{\hbar^2 v_{\rm F}^2
k_{\rm F}^2+\Delta^2}$ is the particle dispersion relation with energy gap
$\Delta=m v_{\rm F}^2$. Moreover, the Fermi velocity, $v_{\rm F}\approx 10^6
$ms$^{-1}$, the Fermi momentum of electron is $k_F$ and $\sigma_i$, where
$i=x,y$ and $z$, are Pauli matrices. We consider a sequence of electron
doped-graphene as wells, and hole-doped graphene as barriers, a schematic of
which with the associated potential is illustrated in Fig. \ref{fig:t1}. The
growth direction is taken to be the $x$ axis, which is designed as the
superlattice axis. The coordinate of the $i$th interface is
labeled by $l_i$ where, $l_i={\rm integer}[\frac{i}{2}]D+integer[\frac{i-1}{2}]
L$. The schematic diagram of the electronic spectrum of the gapped
graphene is shown in Fig. 1 (top graph) as well. Due to the difference between
the Fermi energy and the band structure between two graphene strips, the
potential profile of the system is the multiple quantum well structure which
is described by
\begin{eqnarray}
V(x) =
\left\{%
\begin{array}{ll}
V_0 , & \hbox{ $ {\rm if} \hskip 0.5cm  l_{2i-1}< |x| < l_{2i};$}\\
0 , & \hbox{${\rm otherwise}$}. \\
\end{array}%
\right.
\end{eqnarray}

To solve the transport problem in a graphene
superlattice, we assume that the incident electron propagates at angle $\phi$
along the $x$ axis (see Fig. 1) with energy $E=2\pi v_{\rm F}/\lambda$, and
with the wavelength $\lambda$ across the barriers, in such a way that the
Fermi level lies in the conduction band outside the barrier and the valence
band inside it. Throughout the paper, we consider the Klein zone in which
$\Delta <E< V_0-\Delta$. The Dirac spinor components that are the solutions to
the Dirac Hamiltonian are expressed as
\begin{eqnarray}
\psi_1(x,y)&=& (a_i e^{i k_{ix} x}+b_i e^{-i k_{ix}
x})e^{i k_{y} y}\nonumber\\  \psi_2(x,y)&=&s_i(a_i e^{i
k_{ix} x + i \varphi_i}-b_i e^{-i k_{ix} x - i \varphi_i})e^{i k_{y} y}
\end{eqnarray}
where $a_i$ and $b_i$ are the transmission amplitudes. Here $s_i=sgn(E-V(x))$,
$k_{x}^{2}=(E^{2}-\Delta^{2})/\hbar^{2}
v_{\rm F}^{2}-k_{iy}^{2}$, and, $q_{x}^{2}=((E-V_{0})^{2}-\Delta^{2})/\hbar^{2}
v_{\rm F}^{2}-k_{y}^{2}$, with, $k_{ix}$ is being $k_x $ or $q_x $. Moreover,
$\varphi_i$ is either $\phi$ or $\theta$ for the well and the barrier,
respectively. $k_{x}=k_{\rm F}\cos\phi$ and $k_{y}=k_{\rm F}\sin\phi$ are the
wave vector components for the outside region of the barriers.

To calculate the transmission coefficients, we use the transfer-matrix method.
To this end, we apply the continuity of the wave functions at the boundaries
and construct the transfer matrices as follows:
\begin{eqnarray}
 \left(
  \begin{array}{c}
    1 \\
    r \\
  \end{array}
\right)=\frac{1}{\sin(\alpha_{k})\cos(\phi_{i})}M S(x){\cal N} a_{N}
\label{eq:transfer}
\end{eqnarray}
where
\begin{eqnarray}
M=\left( \begin{array}{cc} \rho_{2}\eta_{1} e^{-i\phi}-\rho_{1}\eta_{2}e^{i\theta} & \rho_{2}\eta_{1}e^{-i\phi}-\rho_{1}\eta_{2}e^{-i\theta} \\
                              \rho_{2}\eta_{1} e^{i\phi}-\rho_{1}\eta_{2}e^{i\theta} & \rho_{2}\eta_{1}e^{i\phi}-\rho_{1}\eta_{2}e^{-i\theta} \\
                             \end{array}
                           \right),  {\cal N}=\nonumber
                                            \left(
                                              \begin{array}{c}
                                                e^{i k_x l_N}(\frac{\rho_{1}}{\rho_{2}}e^{-i\theta}-\frac{\eta_{1}}{\eta_{2}}e^{i\phi})/[2e^{i q_x l_N}\cos \theta] \\
                                                e^{i k_x l_N}(\frac{\rho_{1}}{\rho_{2}}e^{i\theta}+\frac{\eta_{1}}{\eta_{2}}e^{i\phi})/[-2e^{i q_x l_N}\cos \theta] \\
                                              \end{array}
                                            \right)
\end{eqnarray}
\begin{eqnarray}
s(x=l_{i})=\left(
\begin{array}{cc}
t_{11} & t_{12} \\
t_{21} & t_{22} \\
\end{array}
\right)
  ,                        S(x)=s(l_{2}) s(l_{3})...s(l_{N-1})
\end{eqnarray}
and $r$ and $a_N$ are the reflection and the transmission coefficients of the
system that consists of $N$ barriers. We have defined parameters $\rho_{1}=
\cos(\alpha_{k}/2)$, $\eta_{1}=\sin(\alpha_{k}/2)$, $\rho_{2}=
\sin(\acute{\alpha}_k/2)$, and, $\eta_{2}=\cos(\acute{\alpha}_k/2)$. The angles
$\alpha_k$ and $\acute\alpha_k$ are determined by $\tan(\alpha_{k})=\hbar
v_{\rm F}(k_{x}^{2}+k_{y}^2)^\frac{1}{2}/\Delta$ and $\tan(\acute\alpha_{k})=
\hbar v_{\rm F}(q_{x}^{2}+k_{y}^2)^\frac{1}{2}/\Delta$, respectively. The
elements of $s$ matrix have the form
\begin{eqnarray}
t_{11}&=&e^{i({k_{ix}-k_{(i-1)}x)l_{i-1}}}[-\frac{\eta_{i}}{\eta_{i-1}}
e^{i\varphi_{i}}+\frac{\rho_{i}}{\rho_{i-1}} e^{-i \varphi_{i-1}}]
/2\cos(\varphi_{i-1})\\\nonumber
t_{12}&=&e^{i({-k_{ix}-k_{(i-1)}x)l_{i-1}}}[\frac{\eta_{i}}{\eta_{i-1}}
e^{-i\varphi_{i}}+\frac{\rho_{i}}{\rho_{i-1}} e^{-i
\varphi_{i-1}}]/2\cos(\varphi_{i-1})\\
\nonumber
t_{21}&=&e^{i({k_{ix}+k_{(i-1)}x)l_{i-1}}}[-\frac{\eta_{i}}{\eta_{i-1}}
e^{i\varphi_{i}}+\frac{\rho_{i}}{\rho_{i-1}} e^{i
\varphi_{i-1}}]/2\cos(\varphi_{i-1})\\ \nonumber
t_{22}&=&e^{i({-k_{ix}+k_{(i-1)}x)l_{i-1}}}[-\frac{\eta_{i}}{\eta_{i-1}}
e^{-i\varphi_{i}}+\frac{\rho_{i}}{\rho_{i-1}} e^{i \varphi_{i-1}}]/2\cos(\varphi_{i-1}).\nonumber
\end{eqnarray}

The angle-dependence of the transmission probability $T=|a_{N}|^2$
is obtained by solving Eq. (\ref{eq:transfer}) for a given $N$. It should be
noted that the transmission coefficients for the gapless graphene \cite{nima}
is revealed by setting $\Delta=0$ in Eqs. (2)-(5). After the transmission
coefficients are obtained, the conductivity of the
system is computed by means of the B\"{u}ttiker
formula \cite{datta}, taking the integral of $T(E,\phi)$ over the angle,
\begin{equation}
G = G_0 \int^{\frac{\pi}{2}}_{-\frac{\pi}{2}}T(E,\phi)\cos(\phi)d\phi
\label{eq:dc}
\end{equation}
where $G_0=e^2 m_e v_{\rm F} w/\hbar^2$ with $w$ being the width of the
graphene strip along the $y$ direction.

\subsection{Exact analysis for single- and double-barrier systems}

Let us first consider a system composed a single barrier. The wave functions
in the different regions can be written as
\begin{eqnarray}
\psi_1(x,y)&=& e^{(ixk_{x}+yk_{y})}\left(
                             \begin{array}{cc}
                              \rho _{1} \\
                               e^{i\phi}s\eta_{1} \\
                             \end{array}
                           \right)+re^{i{(-xk_{x}+yk_{y})}}\left(
                             \begin{array}{cc}
                              \rho _{1} \\
                               e^{i(\pi-\phi)}s\eta_{1} \\
                             \end{array}
                           \right)\nonumber\\
                            \psi_2(x,y)&=&ae^{i(xq_{x}+yk_{y})}\left(
                             \begin{array}{cc}
                              \rho_{2} \\
                               e^{i\theta}s{'}\eta_{2} \\
                             \end{array}
                           \right)+be^{-i(xq_{x}+yk_{y})}\left(
                             \begin{array}{cc}
                              \rho _{2} \\
                               e^{i(\pi-\phi)}s'\eta_{2} \\
                 \end{array}
                           \right)\nonumber\\
                           \psi_3(x,y)&=&te^{i(xk_{x}+yk_{y})}\left(
                             \begin{array}{cc}
                              \rho_{1} \\
                               e^{i\theta}s\eta_{1} \\
                             \end{array}
                           \right)
\end{eqnarray}
It should be noted that since the interface is located along the $y$, due to
the conservation of the momentum we have, $k_y=q_y$. After some straightforward
calculations, the electronic transmission probability, $T(\phi)$, through the
barrier is obtained, resulting in
\begin{equation}\label{T}
T(\phi)=\frac{\cos^2\theta\cos^2\phi}{\cos^2\phi
\cos^2\theta \cos^2Dq_{x}+\sin^2Dq_{x}(\sin\phi
\sin\theta+\frac{C}{2B})^2}
\end{equation}
where $C/B=\tan{\alpha_{k}/2}\tan{\alpha_{k'}/2}+\cot{\alpha_{k}/2}
\cot{\alpha_{k'}/2}.$

It is useful to investigate the conditions under which the transmission
probability approaches unity. We find from Eq. (\ref{T}) that when
$Dq_{x}=n\pi$ ($n$ is an integer), the barrier becomes entirely transparent,
and does not depend on the parameter $\phi$. The same condition at the normal
incidence was obtained in Ref. [\onlinecite{dombey}].

For a double-barrier system, the calculation of the transmission would be
difficult. In this case we restrict the calculations to the case, $\phi=0$.
The electronic transmission expression for the double-barrier system at normal
incidence takes the form
\begin{equation}\label{T1}
T(\phi=0)=\frac{64(\rho_1 \eta_1 \rho_2 \eta_2)^4}{A^2 \cos^2(2q_xD)
+2AB\cos(2q_xD)[1+(\cos(2q_xD)-1) \cos(2k_xL)]+P+Q}
\end{equation}
where $A=6(\rho_1 \eta_1 \rho_2
\eta_2)^2+(\eta_1 \rho_2)^4+(\rho_1 \eta_2)^4$, $ B=2(\rho_1
\eta_1 \rho_2 \eta_2)^2-(\eta_1 \rho_2)^4-(\rho_1 \eta_2)^4$, $
C=\rho_1 \eta_1^3 \rho_2^3 \eta_2+\rho_1^3 \eta_1 \rho_2
\eta_2^3$, $P=16C^2 \sin^2(2q_xD)+8BC \sin(2k_xL)
\sin(2q_xD)(\cos(2q_xD)-1)$ and $Q=B^2[1+2\cos(2k_xL)(\cos(2q_xD)-1)+(1-\cos(2q_xD))^2]$.
In the numerical section below, we find some critical gap values for
which $T(\phi=0)=1$, and show that the critical points are in good agreement
with the results calculated by the analytical expressions.

\subsection{The Hartman effect in a gapped graphene }

In this section we study tunneling through the single barrier and
calculate two important tunneling times, the group delay time
$\tau_{g}$, and the dwell time $\tau_{d}$ \cite{Hauge}. The
relationship between the two times was first studied
by Winful \cite{Winful} for a one-dimensional electron gas system. It was shown
that there is a difference between the two times in the conventional electron
gas systems. Using the energy derivative of the transmission
phase shift \cite{Winful}, the group delay time is obtained through
$\tau_{gt}=\hbar d\phi_{0}/dE$, where $\phi_{0}=\phi_{t}+k_{x}D$, and the group
delay time in reflection is given by, $\tau_{gr}=\hbar d\phi_{r}/dE$. Here,
$\phi_i$ ($i=t$ or $r$) denotes the phase angle of the transmission or the
reflection wave function.

For a general asymmetric barrier, $\tau_{gt}$ differs from $\tau_{gr}$, and the
group delay time $\tau_{g}$ is obtained by, $\tau_{g}=|t^2|\tau_{gt}+|r^2|
\tau_{gr}$, whereas for symmetric barriers, $\tau_{g}=\tau_{gt}=\tau_{gr}$.
The dwell time - the time spent by a particle in the barrier - is expressed as,
$\tau_{d} =\int^{D}_{0}|\psi(x)|^2 dx/j_{in}$, where $\psi(x)$ is the
stationary state wave function with energy $E$, with $j_{in}=v_{\rm F}
\cos(\phi)$ being the flux of the incident particles.
According to calculations given in Ref [\onlinecite{Zhenhua}] we have
\begin{equation}
\int^{D}_{0}|\psi(x)|^2 dx=-i\hbar v_{\rm F}[(\psi^{\dag}(r) \sigma_{x}
\partial_E \psi(r))_{x=D} - (\psi^{\dag}(r) \sigma_{x} \partial_E
\psi(r))_{x=0}]
\label{eq:hartman}
\end{equation}
For the system, the wave functions are described by
\begin{eqnarray}
\psi_1(x,y)&=& e^{(ixk_{x}+yk_{y})}\left(
                             \begin{array}{cc}
                              \rho _{1} \\
                               e^{i\phi}s\eta_{1} \\
                             \end{array}
                           \right)+re^{i{(-xk_{x}+yk_{y})}}\left(
                             \begin{array}{cc}
                              \rho _{1} \\
                               e^{i(\pi-\phi)}s\eta_{1} \\
                             \end{array}
                           \right)\nonumber\\
\end{eqnarray}
in front of the barrier and
\begin{eqnarray}
 \psi_3(x,y)&=&te^{i(xk_{x}+yk_{y})}\left(
                             \begin{array}{cc}
                              \rho_{1} \\
                               e^{i\phi}s\eta_{1} \\
                             \end{array}
                           \right)
\end{eqnarray}
for behind the barrier. Therefore, the right-hand side of Eq.
(\ref{eq:hartman}) becomes $2v_{\rm F}\rho_{1} \eta_{1}\cos\phi \{|t|^2
\hbar d\phi_{0}/dE +|r|^2 \hbar d\phi_{r}/dE\}$. Consequently, the relationship
between $\tau_d$ and $\tau_g$ is obtained by
\begin{equation}
\tau_{d}=\frac{\varepsilon_k}{\sqrt{{\varepsilon_k}^2+\Delta^2}}~\tau_{g}
\end{equation}
The $\tau_g$ differs from the $\tau_g$ and they are no longer the same in the
presence of the gap values. Note that the energy of a quasiparticle is,
$\varepsilon_k=\hbar v_{\rm F}|\overrightarrow{k}|$. The dwell time becomes
equal to the group delay time by setting $\Delta=0$. The last result is in
contrast to the result obtained for a conventional 2D electron gas system,
where the dwell time equals the group delay time plus a self interference term,
which comes from the overlap of the incident and reflected waves in front of
the barrier.

\section{Numerical Results and discussion}

We evaluated the electronic transmission probability and conductance in
the gapped graphene through a finite number of potential barriers, as a
function of the gap value introduced in the system. In all of the numerical
calculations we assumed that the wavelength of the incident electron is
$\lambda=50$ nm and $V_0=200$ meV. In all the figures $\Delta$ scales in meV.

We first calculated the transmission probability of the charge carriers
through the graphene structure with a double barrier, $N=2$, with the barrier
width $D=50$ nm. Figure 2 shows the transmission probability of the incident
electrons hitting a graphene superlattice as a function of the angle $\phi$
for several values of the gap values, $\Delta$, with (a) $L=50$ nm
and (b) $L=70$ nm, respectively. The magnitude of $T$ behaves non-monotonically
with the increase of the energy gap value at $\phi=0$. It shows that the Klein
tunneling is no longer applicable when there is a band gap in graphene. It
should be noted that the Kelin tunneling predicts that the
chiral massless carrier can pass through a high electrostatic potential
barrier with probability one, regardless of the height and width of the
barrier at normal incidence. In addition, to verify the dependence on the
well width of the transmission probability, we calculated the electronic
transmission for the several values of the parameter $L$. The results are
depicted in Fig. 2(b).

To verify the behavior of the electronic transmission probability at normal
incidence, the calculated transmission probability as a function of the band
gap value for the several numbers of the potential barriers is shown in Fig.
3(a). The structural parameters are the same as in Fig. 2(a). The results for
a single barrier, $N=1$, show that when $\Delta$ increases the transmission
probability exhibits a minimum at $\Delta\simeq 62$ meV, and then it reaches
unity at the critical value given by, $\Delta_c=82$ meV. The critical gap
values entirely coincide with the results calculated analytically. In the case
of $\phi=0$ we found analytically that, $
\Delta_c=\sqrt{(E-V_{0})^2-(\hbar v_{\rm F})^2(n\pi/D})^2$, which is
supported by the numerical calculations.

In the case of the double barrier, $N=2$, there are three $\Delta_c$ values
with which $T$ becomes exactly unity. For the system that consists of an even
number of the potential barriers, we found that the superlattice is
fully transparent ($T=1$), when the energy gap is $\Delta_c=53.5$ meV.
Note that the critical value of the band gap $\Delta_c$ depends on
the superlattice structural parameters, as we show
numerically in Fig. 3(b), where $L=70$ nm. The number of the maxima at which
the transmission amplitude becomes unity increases with increasing the number
of the potential barriers.

We also studied that how the structural parameters affect the transmission of
the system. Figure 4 presents $T(\phi=0)$ as a function of the well width for
the several values of the barrier width. The number of the potential barriers
is $N=2$, and the value of gap was chosen as $\Delta=53.5$ meV. The numerically
calculated $T(\phi=0)$ is in good agreement with the result obtained
analytically using Eq. (\ref{T1}). Furthermore, the transmission
probability approaches unity only for the specific values, $L=0,25,50,75,
100,\cdots$ at $D=50$ meV. However, when $D=20$ or 80 meV, the transmission
probability is independent of the parameter $L$ at normal incidence angle,
and always approaches unity. the numerical results predict that the
transmission probability at normal incidence angle to be unity for the case
that $2q_xD=2n\pi$, and is independent of $L$. However, when $2q_xD=(2n-1)\pi$,
the transmission probability would be unity if $2k_x L=2m\pi$. Therefore,
the transmission probability depends strongly on the structural parameters in
the gapped graphene superlattice.

Finally, we calculated the electronic conductance as a function of $\Delta$
for the various numbers of the potential barriers. The results are shown in
Fig. 5. Finite-size scaling analysis indicates that $G$ tends to a nonzero
constant at $\Delta=80$ meV. Moreover, there is a domain value of $\Delta$
for which the conductivity attains its maximum value. According to the
above discussions, it is clear that the conductivity of the system depends on
the superlattice structural parameters, such as $L$ and $D$. Importantly, we
would like to stress that the electronic conductance can reach a maximum value
by selecting the proper gap and the barrier width.

\section{Summary}

We evaluated the electronic conductance in gapped graphene with a finite
number of potential barriers. An exact analytic expression was derived for the
electronic transmission probability in a system with a single or two barrier,
and the critical values of the gap at which the transmission probability
equals unity were computed. We showed that the group delay time is not the
same as the dwell time in a gapped graphene that consists of a barrier.
However, they are the same in a gapless graphene. It should be noted that the
extension of the dwell time for a superlattice structure needs intensive
computations, and will be reported in future. Moreover, we showed that the
conductance can attain its maximum for a domain value of gaps around the
critical value. In addition, the conductance of the system depends on the
superlattice structural parameters and, therefore, one may design a very good
electronic device by selecting the proper gap and the barrier width. Thus, the
system with a proper arrangement might be of use in electronic or
electromagnetic devices. Finally, the present calculations may be improved to
investigate the spin dependence of the conductance.

\begin{acknowledgments}
We are grateful to M. Sahimi who carefully read the manuscript. A.E. acknowledges supporting from the SRU.
\end{acknowledgments}

\begin{figure}[ht]
\begin{center}
\includegraphics[width=8 cm]{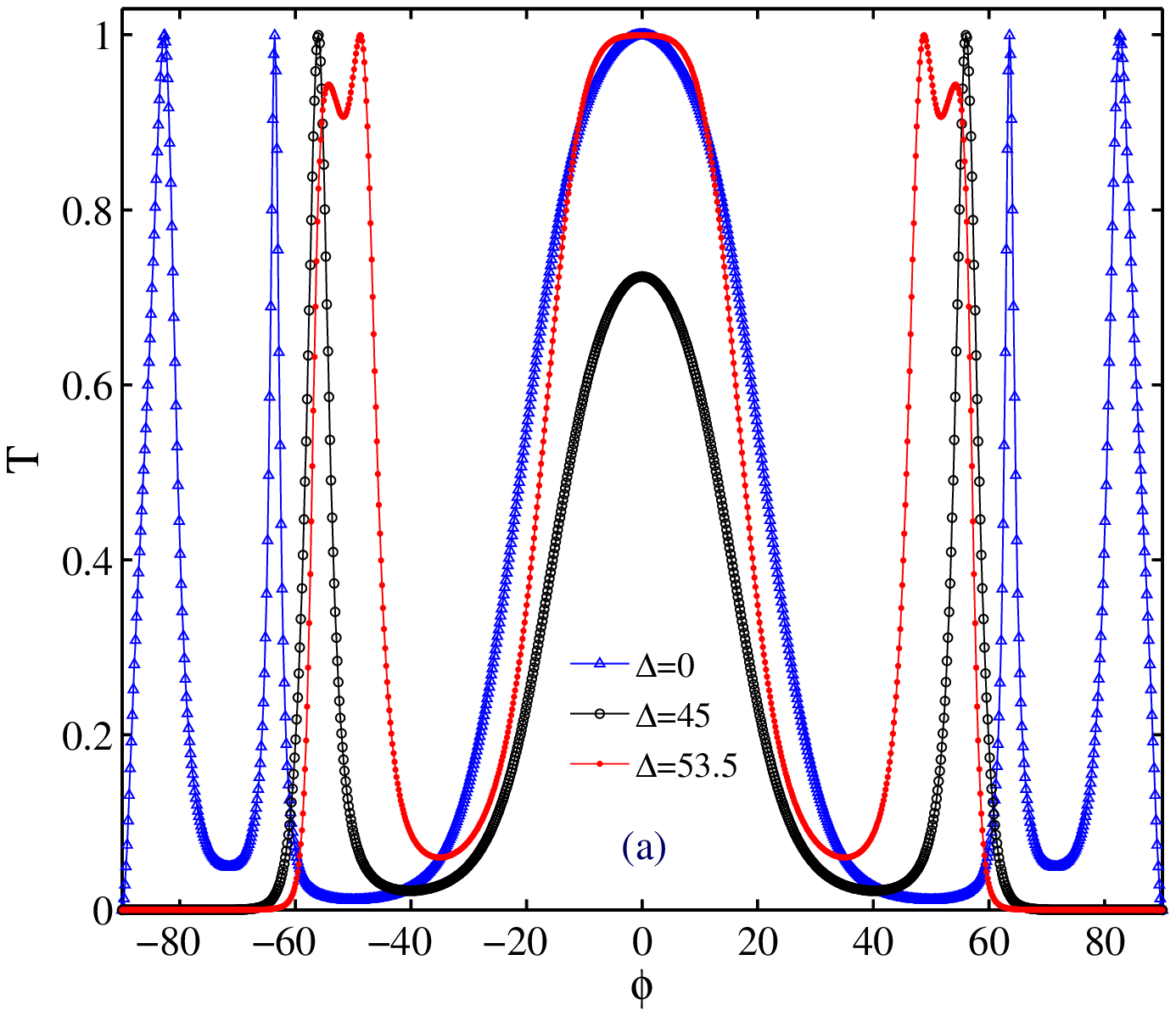}
\includegraphics[width=8 cm]{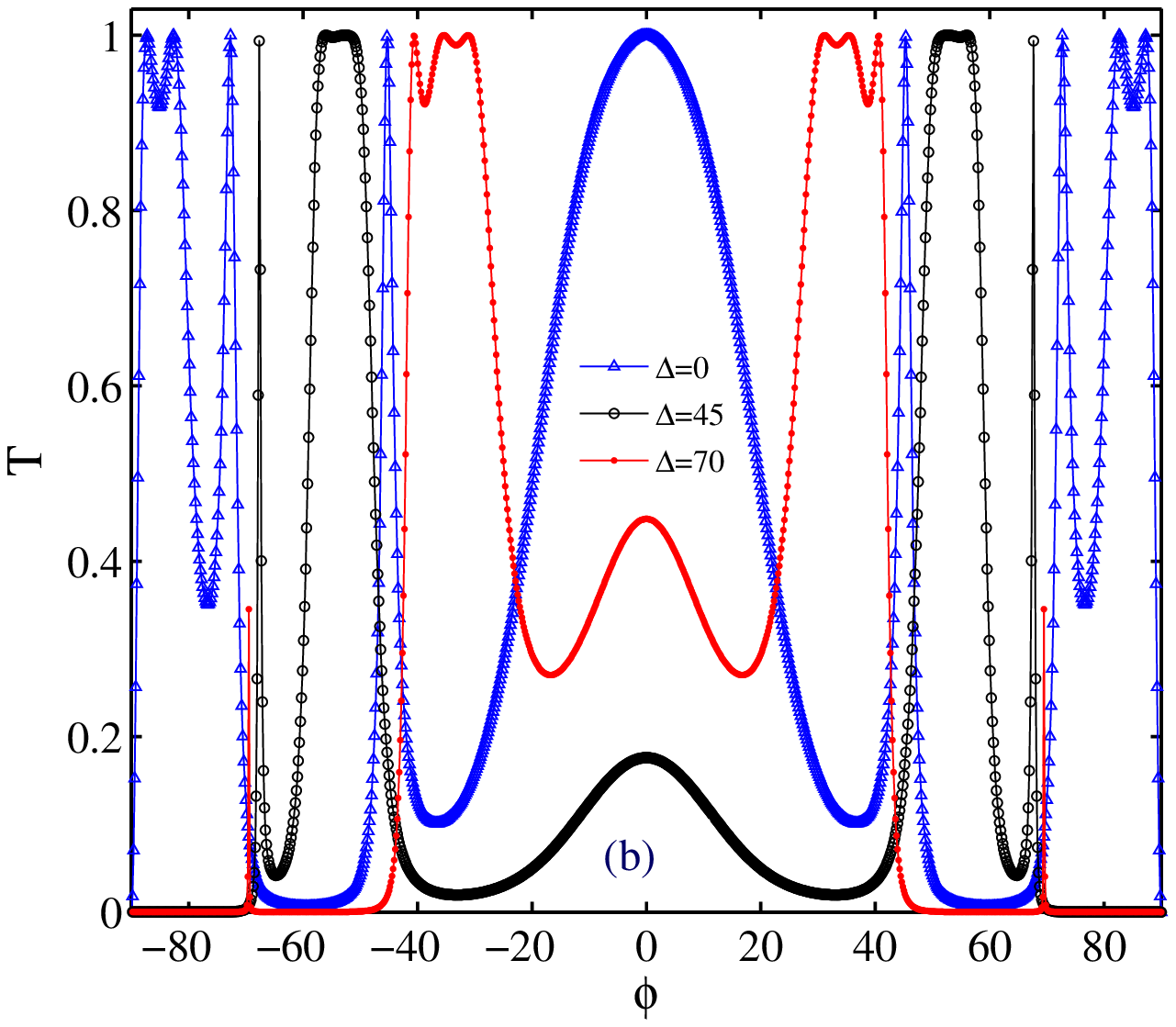}
\caption{(Color online) Transmission probability $T$ of the electrons
through the double-barrier structures as a function of
the incident angle and the parameter $\Delta$. The value of the barrier width
is $D=50$ nm and that of the well width are (a) $L=50$ nm, and (b) $70$ nm.}
 \label{fig:t2}
\end{center}
\end{figure}

\begin{figure}[ht]
\begin{center}
\includegraphics[width=8 cm]{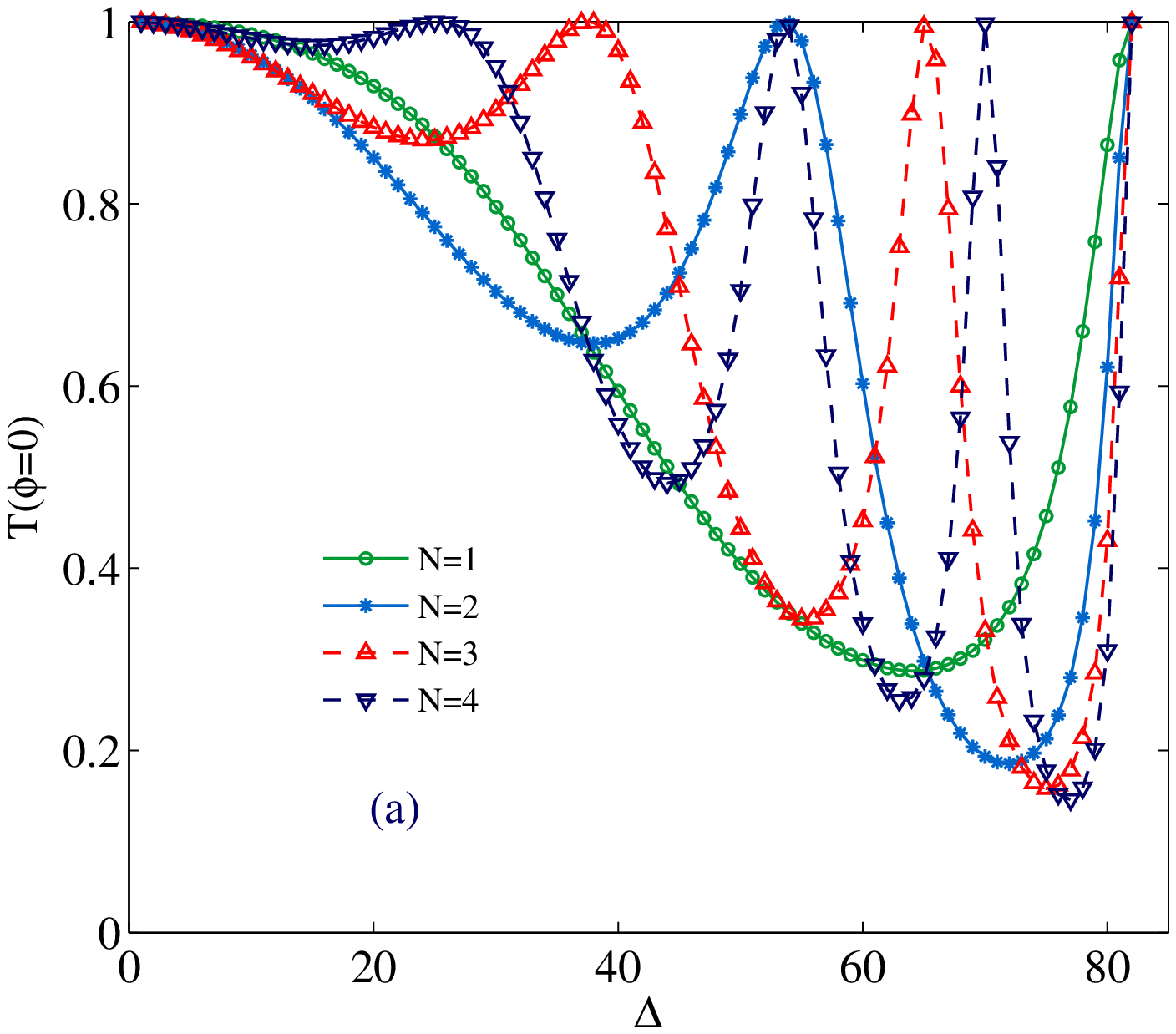}
\includegraphics[width=8 cm]{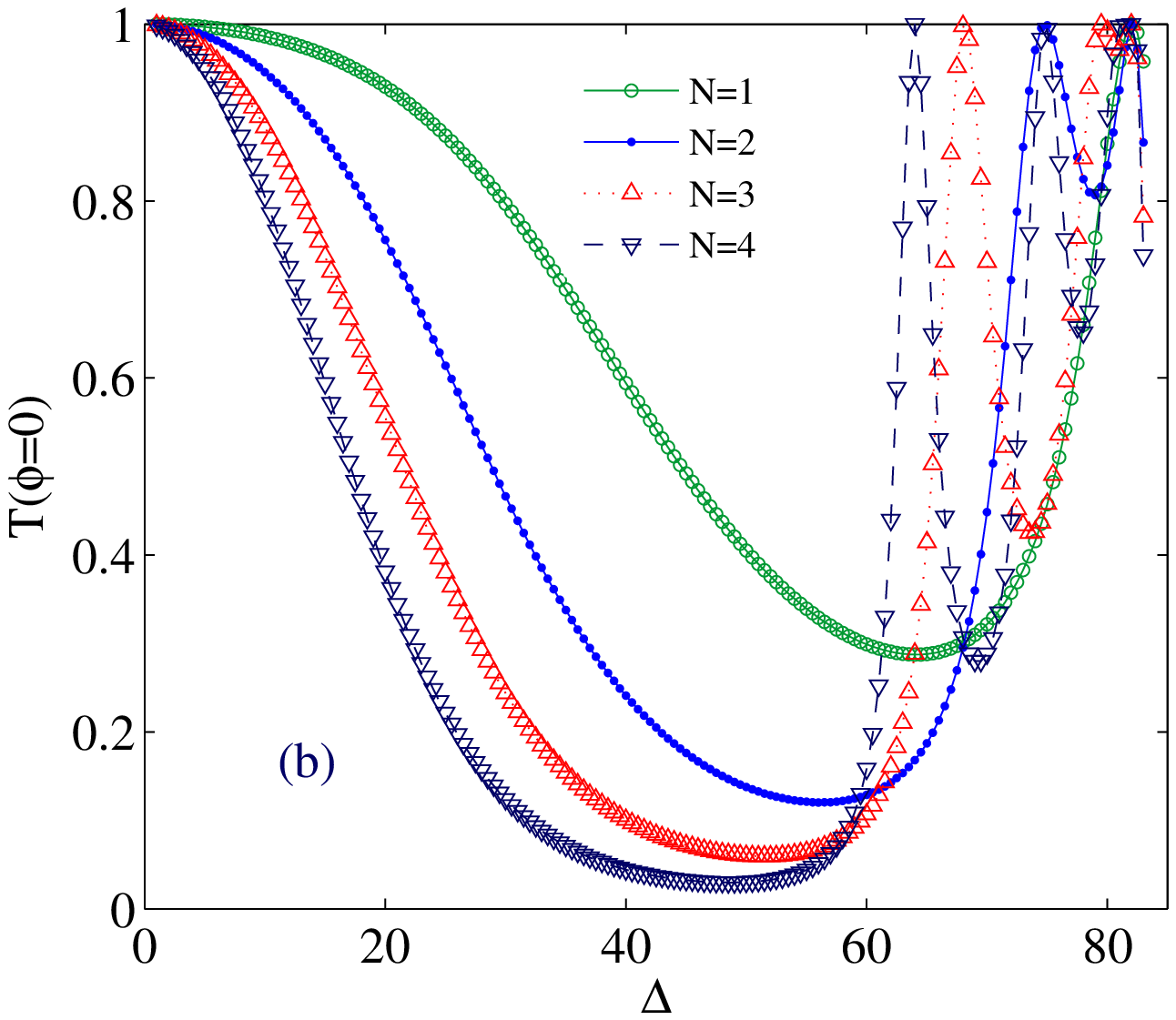}
\end{center}
\caption{Transmission probability $T$ for the normal incident electrons
through graphene superlattice as a function of $\Delta$ for (a) $L=50$ nm, and
(b) $L=70$ nm for several numbers of the barriers. }
\label{fig:t3}
\end{figure}

\begin{figure}[ht]
\begin{center}
\includegraphics[width=8 cm]{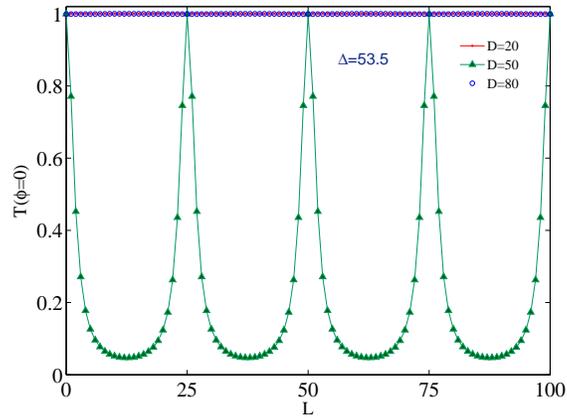}
\end{center}
\caption{Transmission probability $T$ for the normal incident electrons
through the graphene superlattice that consist of $N=2$ as a function of
well's width $L$ at $\Delta=53.5$ meV for several values of the barrier width
$D$. }
\label{fig:t4}
\end{figure}

\begin{figure}[ht]
\begin{center}
\includegraphics[width=8 cm]{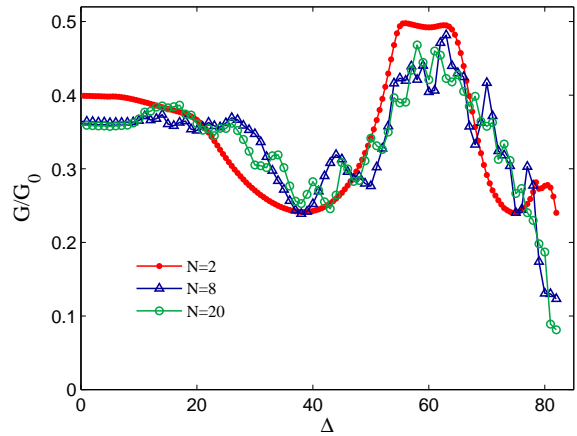}
\end{center}
\caption{Conductance of the graphene superlattice as a function of the
parameter $\Delta$ for $L=50$ nm and several numbers of the barriers. }
\label{fig:t5}
\end{figure}

\end{document}